\pdfoutput=1

\documentclass[reprint,aps,prd,superscriptaddress,showpacs,preprintnumbers,amsmath,amssymb,floatfix]{revtex4-1} 

\usepackage[mathlines]{lineno}
\usepackage{bm}
\usepackage{color}
\usepackage{upgreek}
\usepackage{afterpage}
\usepackage{tikz, pgfplots}
\usepackage{mhchem} 
\usepackage[colorlinks=true]{hyperref}

\newcommand{\etal}{\textit{et al}.}

\newcommand{\fnr}{\ensuremath{F_{\mathrm{nr}}}} 
\newcommand{\signrEDW}{\ensuremath{\sigma^{\mathrm{EDW}}_{\mathrm{NR}}}} 
\newcommand{\signr}{\ensuremath{\sigma_{\mathrm{NR}}}} 
\newcommand{\sigerEDW}{\ensuremath{\sigma^{\mathrm{EDW}}_{\mathrm{ER}}}} 
\newcommand{\siger}{\ensuremath{\sigma_{\mathrm{ER}}}} 
\newcommand{\signrI}{\ensuremath{\sigma^{\mathrm{SAI}}_{\mathrm{NR}}}} 
\newcommand{\signrIst}{\ensuremath{\sigma^{\mathrm{SA}}_{\mathrm{NR}}}} 

\begin{document}

\preprint{ \today}

\title{The intrinsic Fano factor of nuclear recoils for dark matter searches
}

\author{M.~Matheny} \affiliation{Department of Physics, University of Colorado Denver, Denver, Colorado 80217, USA} 
\author{A.~Roberts} \email{Corresponding author: amy.roberts@ucdenver.edu} \affiliation{Department of Physics, University of Colorado Denver, Denver, Colorado 80217, USA} 
\author{A.~Srinivasan} \affiliation{Department of Physics, University of Colorado Denver, Denver, Colorado 80217, USA} 
\author{A.N.~Villano} \email{Corresponding author: anthony.villano@ucdenver.edu} \affiliation{Department of Physics, University of Colorado Denver, Denver, Colorado 80217, USA}

\smallskip
\date{\today}

\noaffiliation


\smallskip

\begin{abstract}
Nuclear recoils in germanium and silicon are shown to have much larger variance in electron-hole
production than their electron-recoil counterparts for recoil energies between 10 and 200\,keV.
This effect--owing primarily to deviations in the amount of energy given to the crystal lattice in
response to a nuclear recoil of a given energy--has been predicted by the Lindhard model.  We
parameterize the variance in terms of an intrinsic nuclear recoil Fano factor which is
24.3$\pm$0.2 and 26$\pm$8 at around 25\,keV for silicon and germanium respectively. The variance
has important effects on the expected signal shapes for experiments utilizing low-energy nuclear
recoils such as direct dark matter searches and coherent neutrino-nucleus scattering measurements.  

\end{abstract}


\maketitle

%
%
%
%
%
%
%

%
%
%
%
%
%
%
%

\section{\label{sec:intro}Introduction}
One of the most intensely researched channels for direct detection of dark matter is scattering
off of a nucleus in a target
material~\cite{ARNAUD201854,PhysRevD.100.102002,Angloher2014,PhysRevLett.121.111302,PhysRevLett.120.061802}.
While the ionization distribution for these recoils has never been well understood in solids, the
Lindhard model~\cite{osti_4701226} provides a benchmark.  Experiments that simultaneously measured
two deposition channels (like ionization and heat) did not worry about the exact ionization
distributions in the past because they could measure recoil energy
directly~\cite{PhysRevD.99.082003,PhysRevD.92.072003} (without a model for ionization production).
With the two-channel measurements (ionization and heat) the results of dark matter searches was
not systematically limited due to the lack of knowledge on the ionization.

In recent years, there has been a dramatic improvement in the detection energy threshold of many
experiments~\cite{PhysRevD.99.062001,PhysRevLett.125.241803}, due largely to improvements in the
measurement resolution for ionization or heat individually.  The best detectors of the new
generation of low-mass dark-matter-seeking experiments have single electron-hole pair
sensitivity~\cite{PhysRevLett.119.131802,doi:10.1063/1.5010699}.  These detectors have not yet
been able to achieve the ionization-yield insensitivity that their higher-energy predecessors
have, so the dark-matter signal depends sensitively on the ionization yield and ionization
variance produced by a low-energy nuclear recoil. In fact, it is often true that dominant
systematic uncertainties in dark matter limits come from the uncertainty in the ionization
yield~\cite{PhysRevD.99.062001}. For single electron-hole devices the ionization variance also
becomes a driving factor in the accuracy of signal models for low-mass dark matter via nuclear
scattering.  

While much of the literature has focused on the ionization
yield~\cite{PhysRevD.105.083014,PhysRevD.94.122003,PhysRevD.94.082007,PhysRevD.91.083509}, there are existing
published data that constrain the ionization variance either directly or
indirectly~\cite{PhysRevA.45.2104,PhysRevD.42.3211,MARTINEAU2004426}. And there is even more data
still that might be used to more precisely measure the ionization variance if a resolution model
was published~\cite{PhysRevLett.112.241302}. 

We report here on the best such existing data to constrain the ionization variance in silicon and
germanium, and provide a procedure by which such information can be extracted from a dark matter
detector that measures two channels like ionization and heat. While our constraints are limited to
the recoil energy region above about 24\,keV, the techniques give insight into how this information
can be extracted to lower energies in the future and the basic size and trend of the ionization
variance for nuclear recoils.

To analyze the silicon data we have taken note that two previous publications have reported an
``excess'' ionization variance beyond the expected instrumental variance. We converted that
variance into a Fano factor by using $\sigma_e^2$/$\bar{N}$=$F$, where $\sigma_e$ is the excess
width in the ionization measurement. 

We have used a similar method for a previous germanium publication. There, we used electron
recoils to constrain the instrumental resolutions, obtaining an excellent fit to the measured
widths for electron recoils (see Fig.~\ref{fig:edw_ERQ}). This instrumental resolution does not
predict the nuclear recoil widths, so we include an ``excess'' variance in the form of a nuclear
recoil Fano factor to obtain good fits. The method is similar to that of scintillator
references~\cite{BOUSSELHAM2010359} who have accounted for instrumental resolution in order to
obtain information on the intrinsic optical photon production process. The key difference being
that the result in that publication shows sub-Poisson fluctuations in optical photon production
where we see larger-than-Poisson fluctuations in electron-hole pair production for nuclear recoils. 

Our results for the intrinsic nuclear recoil Fano factors of silicon and germanium (see
Figs.\ref{fig:silicon_effF} and~\ref{fig:germanium_effF}) are not in line with a division of some theoretical predictions of the electron recoil Fano
factor~\cite{Mei_2020} by the ionization yield; this naive modification gives $F\sim$0.41. Those predictions, however, assume an underlying
phonon distribution that is Poisson. Our results are roughly in line with the
Lindhard~\cite{osti_4701226} predictions which do not make that assumption.

\section{\label{sec:si_fano}The Fano factor}

The ionization variance for electron recoils is very succinctly characterized in terms of the Fano
factor, F~\cite{PhysRev.70.44}. Given an average number of electron-hole pairs produced,
$\bar{N}$, the variance in this number of pairs is given simply by:

\begin{equation}
	\sigma_{N}^2 = F\bar{N}. 
\end{equation}

While this specification does not give insight into moments of the $N$ distribution of higher
order than the variance, it emphasizes that $F$=1 corresponds to a behavior that is
\emph{qualitatively} similar to a Poisson distribution in the lowest two moments. For these reasons we
find it simple and convenient to parameterize the nuclear recoil ionization variance in the same
way, but with a modified intrinsic Fano factor, \fnr~\footnote{Note that the convenience is in
comparison to the electron recoil case and the qualitative similarity to the Poisson distribution, the
underlying causes of the variance are different in the nuclear recoil case, arising from the
inherent randomness in the energy deposition process in terms of the fraction of energy given to
the lattice. The electron recoils have approximately zero energy given to the lattice.}. The Fano
factor for electron recoils seems to be in the range\footnote{The measurement temperatures are
different for these measurements, being 90\,K, 77\,K, and 87\,K for the lowest-to-highest
measurements} 0.084\,--\,0.16~\cite{YAMAYA1979181,EBERHARDT1970291,PALMS196959}, but may have a
temperature and/or energy dependence~\cite{PEROTTI1999356}. It is even possible that the ``true''
intrinsic Fano factor has not yet been measured directly and is lower than all the above
measurements~\cite{4325691}. 

In silicon there are two studies that we are aware of that measured the ionization variance in
addition to the ionization yield for nuclear recoils. Both were done in the early 1990's with
secondary neutron beams produced from primary proton beams via the reaction
$^7$Li(p,n)$^7$Be~\cite{PhysRevA.45.2104,PhysRevD.42.3211}. The measurement by Dougherty makes use
of neutron elastic-scattering resonances present in silicon. The Gerbier~\etal~measurement
uses a fixed-angle secondary neutron detector and a timing coincidence to constrain the true
recoil energy in the silicon scattering detector.

Both of these measurements report the ``extra'' ionization variance after subtracting the expected
variance due to known sources of errors such as instrumental noise or angular uncertainty in the
secondary neutron scatters. The extracted additional ionization variance can be compared with the
total recoil energy (inferred in the Dougherty measurement and measured in the Gerbier
measurement) to give what we will define as the intrinsic fractional ionization width, $\xi$.
This fractional ionization width is defined as the ionization width (in energy units)
divided by the ionization energy collected, so that $\xi = \sigma_N/\bar{N}$. With these
definitions the intrinsic nuclear-recoil Fano factor, \fnr, is given by:

\begin{equation}\label{eq:eff_fano}
	\fnr = \bar{N}\xi^2 = \frac{E_r\bar{Q}}{\epsilon_{\gamma}} \xi^2,
\end{equation}

where $E_r$ is the true recoil energy, $\bar{Q}$ is the average ionization yield (ratio
of ``collected'' ionization energy to total energy; unity for electron recoils), and
$\epsilon_{\gamma}$ is the average energy to produce one electron-hole pair for an electron
recoil. 

\begin{table*}[!hbt]
\begin{tabular}{ c  c  c  c  c  c }
\hline
\hline
Recoil Energy (keV)  &  Ion. Efficiency (\%)   & Non-instr. Width (eV) & Non-instr. Width (\%) & Effective Fano & Reference  \\ \hline
	109.1$\pm$0.7 & 51.4$\pm$2 & - & 6.1$\pm$1.2 & 208$\pm$82 & Dougherty~\cite{PhysRevA.45.2104} \\
	75.7 $\pm$0.4 & 45.6$\pm$0.5 & - & 5.3$\pm$0.6 & 123$\pm$28 & Dougherty~\cite{PhysRevA.45.2104} \\
	25.3$\pm$0.3 & 35.5$\pm$0.6 & - & 3.6$\pm$0.3 & 24.3$\pm$4.1 & Dougherty~\cite{PhysRevA.45.2104} \\
	7.50$\pm$0.03 & 26.9$\pm$0.4 & - & 2.8$\pm$0.4 & 5.75$\pm$1.65 & Dougherty~\cite{PhysRevA.45.2104} \\
	4.15$\pm$0.15 & 22.5$\pm$0.5 & - & 2.2$\pm$0.9 & 2.35$\pm$1.92 & Dougherty~\cite{PhysRevA.45.2104} \\
	21.7$\pm$0.2 & 40.7$\pm$0.5 & 1000$\pm$59 & - & 29.80$\pm$17.11 & Gerbier~\cite{PhysRevD.42.3211} \\
	19.5$\pm$0.2 & 38.7$\pm$0.7 & 1101$\pm$108 & - & 42.27$\pm$8.33 & Gerbier~\cite{PhysRevD.42.3211} \\
	13.5$\pm$0.3 & 33.6$\pm$0.7 & 601$\pm$42 & - & 20.96$\pm$2.96 & Gerbier~\cite{PhysRevD.42.3211} \\
	8.6$\pm$0.1 & 31.1$\pm$0.5 & 348$\pm$13 & - & 11.91$\pm$0.91 & Gerbier~\cite{PhysRevD.42.3211} \\
	4.7$\pm$0.1 & 26.6$\pm$0.8 & 185$\pm$36 & - & 7.20$\pm$2.81 & Gerbier~\cite{PhysRevD.42.3211} \\
	4.15$\pm$0.1 & 27.4$\pm$0.8 & 166$\pm$39 & - & 6.38$\pm$3.00 & Gerbier~\cite{PhysRevD.42.3211} \\
	3.9$\pm$0.1 & 22.9$\pm$2.0 & 241$\pm$66 & - & 17.11$\pm$9.49 & Gerbier~\cite{PhysRevD.42.3211} \\
	3.3$\pm$0.1 & 25.9$\pm$1.6 & 131$\pm$55 & - & 5.28$\pm$4.45 & Gerbier~\cite{PhysRevD.42.3211} \\

\hline
\hline
\end{tabular}
   \caption{\label{tab:si_data}Data from the past publications constraining the intrinsic Fano
	factor in silicon~\cite{PhysRevA.45.2104,PhysRevD.42.3211}. The intrinsic Fano factor is
	calculated from Eq.~\ref{eq:eff_fano} in the text, using the measurements of the
	non-instrumental widths and the ionization efficiency (yield).  
   }
\end{table*}

Table~\ref{tab:si_data} shows the resulting intrinsic nuclear-recoil Fano factors for the silicon
nuclear recoils measured in the two references we have been discussing. Even at low recoil
energies, around 3\,keV, the intrinsic Fano factors show that the ionization variance is such that
the number of created pairs have more variance than a Poisson process with the same average number
of pairs. 

The Lindhard~\etal~model, articulated in the early
papers~\cite{lindhard1968approximation,osti_4153115,osti_4701226,PhysRev.124.128}, contains
predictions for the variance in the production of electron-hole pairs in a solid medium in
addition to the \emph{average} ionization (ionization yield). We compare this theoretical
ionization variance with ``extra'' ionization variance extracted by Dougherty and Gerbier as a
possible explanation. 

\begin{figure}[!htb]
    \includegraphics[width=\columnwidth]{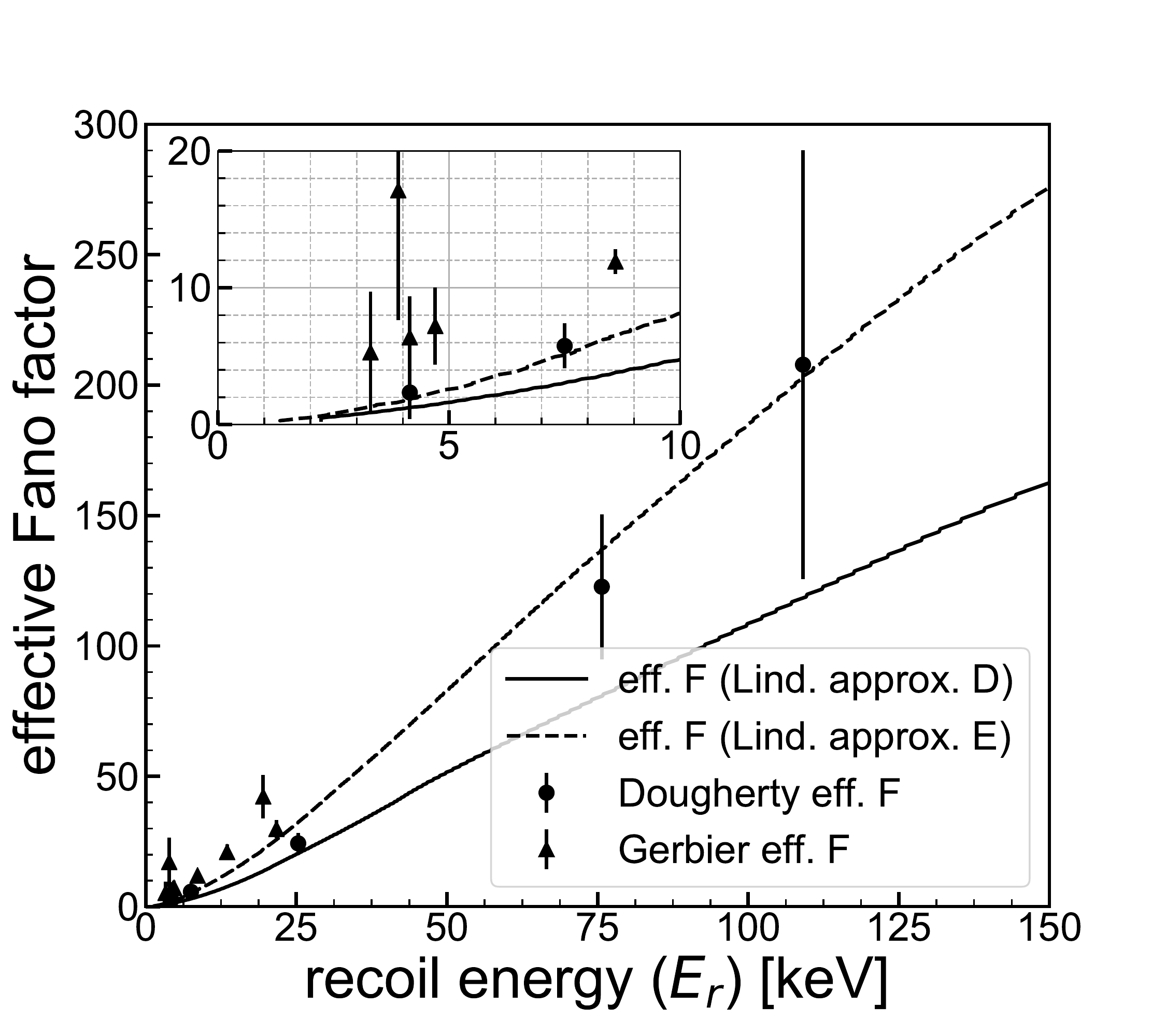}
	\caption{\label{fig:silicon_effF} The measurements of Dougherty~\cite{PhysRevA.45.2104}
	and Gerbier~\cite{PhysRevD.42.3211} converted into the intrinsic Fano factor for nuclear
	recoils. We also show the predictions of Lindhard~\cite{osti_4701226} in the so-called
	Approximation D (solid) and Approximation E (dashed) curves (see text for descriptions).
	The inset shows a zoom of the low-energy region below 10\,keV in recoil energy.  
	}
\end{figure}

Figure~\ref{fig:silicon_effF} shows the ionization variance results of the previous measurements
by Dougherty and Gerbier cast in terms of the intrinsic nuclear recoil Fano factor. Lindhard's
predictions--shown in terms of the intrinsic Fano factor--are also shown on the plot. The
predictions shown employ two different approximations used in the Lindhard work: the approximate
separability between electronic and nuclear energy deposits (referred to as approximation D); and
the additional assumption of forward-scattering dominance in nuclear collisions (referred to as
Approximation E). Approximation E produces a lower ionization yield and a larger ionization
variance. The lower ionization yield for Approximation E is expected because in that case nuclear
collisions are assumed to transfer small amounts of energy and therefore contribute a smaller
fraction of their energy to ionization. Furthermore, the larger ionization variance is a
consequence of the approximate proportionality between \fnr\,and $\sqrt{\epsilon}$, where
$\epsilon$ is the average energy needed to create a single electron-hole pair~\cite{Mei_2020}--a
quantity that increases with decreasing yield. 

Despite clear evidence for a very large ionization production variance for nuclear recoils, and
the importance of this variance for low-mass dark matter searches, studies of this effect are
scant. Dark matter collaborations like SuperCDMS and EDELWEISS have excellent sensitivity to this
effect because of their direct measurements of ionization yield. In the next sections we argue
that the large ionization variance expected in moderate-energy nuclear recoils produces
larger-than-expected measured ionization yield widths in cryogenic semiconductor detectors, and
that this fact can be used to measure the ionization variance for silicon or germanium.  

\section{\label{sec:edw_yield}Previous germanium ionization yield measurement}
While the previously discussed measurements of the ionization variance in silicon came in the
early 1990's, other technologies that came later had excellent means to probe the ionization
variance in germanium. Two such similar technologies came out of the cryogenic dark matter
searches of EDELWEISS and SuperCDMS~\cite{MARTINEAU2004426,PhysRevLett.112.241302}. 

EDELWEISS~\cite{MARTINEAU2004426} was possibly the first to note in published work that the
nuclear-recoil band in cryogenic ionization/phonon devices is expected to be significantly
\emph{narrower} than the electron-recoil band if only the effects of sensor resolution are
included. Recently, the narrowness of the nuclear-recoil band when using empirical resolution
functions has also been noted in the SuperCDMS detectors~\cite{AllisonThesis}. The width of the
nuclear-recoil band is directly related to the variance of the ionization yield (or what EDELWEISS
and some others call the ``Quenching'').  In this work we use $Q$ to denote the random variable
corresponding to the \emph{measured} ionization yield for an event, and $\bar{Q}$ to denote the
average of that quantity, equivalent to the $\langle Q\rangle$ of EDELWEISS.

At a given recoil energy the width of the quenching measurement was estimated in the 2004 EDELWEISS
work~\cite{MARTINEAU2004426} by~\footnote{Actually because of the definition of quenching as
$E_I/E_r$ the expected width on this quantity needs to be derived more carefully than the standard
$\sqrt{(df/dA)^2\sigma_A^2 + (df/dB)^2\sigma_B^2}$.}:

\begin{equation}\label{eq:sigma_Q_EDW}
	\left(\signrEDW\right)^2 = \frac{1}{E_r^2}\left ((1+\frac{V}{\epsilon_{\gamma}}\bar{Q})^2 \sigma_I^2 \\ 
	+ (1+\frac{V}{\epsilon_{\gamma}})^2\bar{Q}^2 \sigma_H^2 \right),
\end{equation}

where $\sigma_I^2$ is the variance in the ionization signal in energy units and $\sigma_H^2$ is
the variance in the heat signal in energy units. Since the quenching factor (less than unity for
nuclear recoils) decreases each term in the equation, it is easy to see the variance in the
event-by-event measured quenching should be significantly less for nuclear recoils than for
electron recoils. In fact, this is not the case. EDELWEISS measures the variance in the nuclear
recoils to be comparable to that of the electron recoils~\cite{MARTINEAU2004426}. We have
reproduced the EDELWEISS analysis by first computing the expected ionization yield width for
electron recoils and then doing a simple fit to constrain how much larger the nuclear recoil width
that EDELWEISS measures is from the prediction in Eq.~\ref{eq:sigma_Q_EDW}.

For the electron recoils, the average ionization yield, $\bar{Q}$, is taken to be unity. EDELWEISS
parameterized the energy-dependent sensor resolutions by the following functional
forms~\cite{MARTINEAU2004426}: 
\begin{equation}\label{eq:edw_res}
\begin{aligned}
	\sigma_I(E) &= \sqrt{(\sigma_I^0)^2 + (a_IE)^2} \\
	\sigma_H(E) &= \sqrt{(\sigma_H^0)^2 + (a_HE)^2},
\end{aligned}
\end{equation}

where $a_I$ and $a_H$ are adjustable parameters used to ``tune'' the ionization yield width.
Using these resolutions we can compute the \emph{exact} ionization yield width as a function of
energy~\footnote{As noted before the electron recoils also have a Fano factor but its effect was
included in the measurement of the EDELWEISS resolution parameters.}.  The resolutions can also be
used to calculate the nuclear-recoil width given an intrinsic Fano factor for nuclear recoils. The
calculations are outlined in the Appendix. 

\begin{figure}[!h]
      \includegraphics[width=\columnwidth]{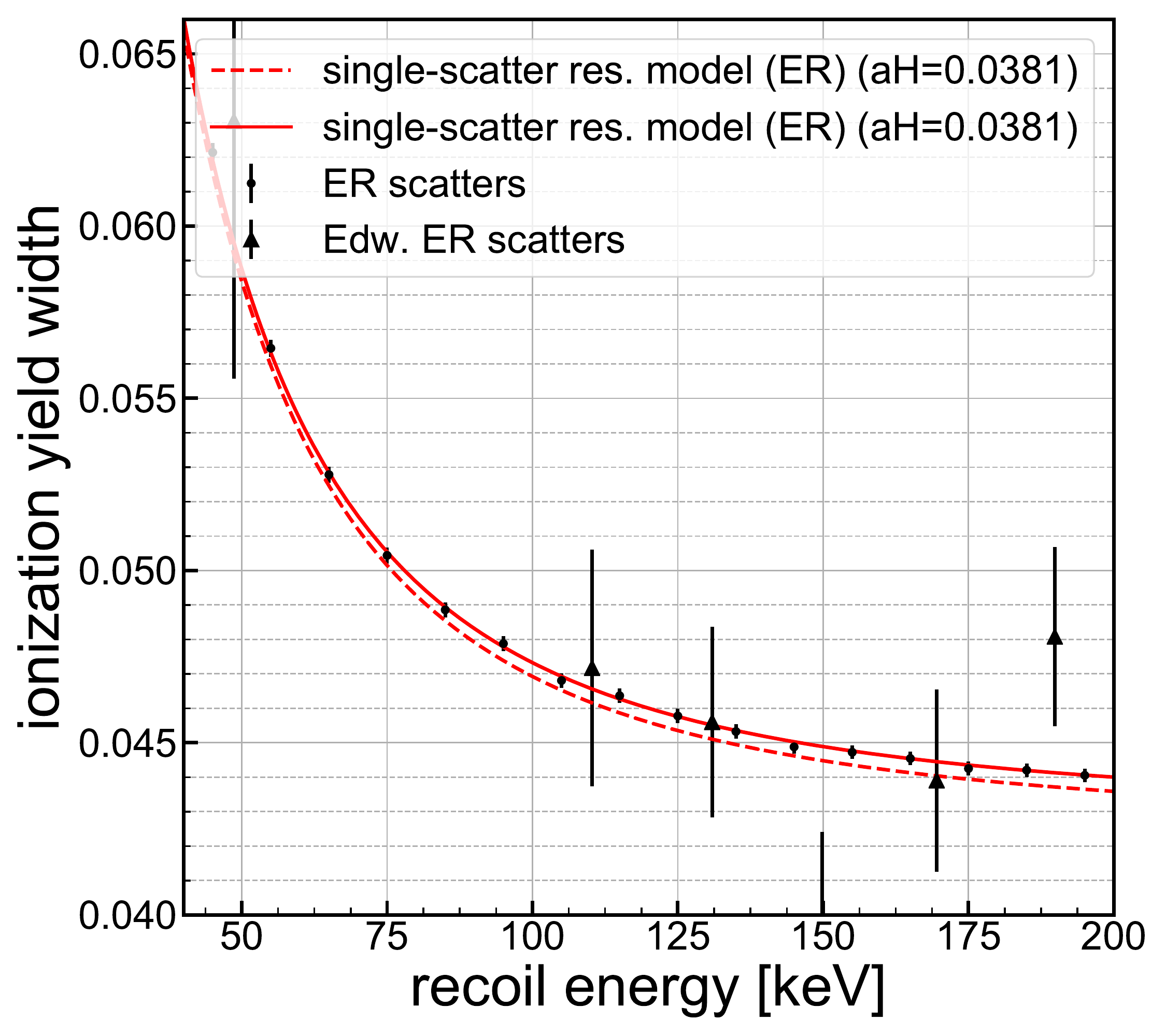}

	\caption{\label{fig:edw_ERQ} (Color online) The measured ionization width for
	electron recoils. The triangular data points are from the
	EDELWEISS~\cite{MARTINEAU2004426} measurement on detector GGA3, and the circular data
	points are our simulation of that measurement with the ``tuned'' resolutions (see text).
	The solid curve is our exact model (\siger) for the ionization width given the appropriate
	resolution and the dashed curve is the zeroth-order model (\sigerEDW) used by EDELWEISS.
	}
\end{figure}

Figure~\ref{fig:edw_ERQ} shows the energy dependence of our exact expression for the
electron-recoil width (see the Appendix) with the EDELWEISS approximations
(Eq.~\ref{eq:sigma_Q_EDW}).  Looking at the electron-recoil band of the EDELWEISS GGA3 detector,
we tune the parameter $a_H$ to be 0.0381 in FWHM units~\footnote{Eq.~\ref{eq:edw_res} has the same
form when written in terms of the instrumental FWHM values, except that the constants will have to
be multiplied by a factor, due to convention $a_H$ is reported as if this equation is the FWHM
version}. This value gives the best fit to the EDELWEISS electron-recoil yield widths as a
function of energy, as shown in Fig.~\ref{fig:edw_ERQ}. Also in the figure we have displayed the
widths resulting from a simulation of these distributions given the tuned sensor resolutions and
our best expression for the electron-recoil yield width given in the Appendix.  We see that the EDELWEISS
approximation to the yield width (Eq.~\ref{eq:sigma_Q_EDW}) is lower by an amount that seems
unimportant for this analysis, given the precision of the electron-recoil yield width data, but
the exact expression (see the Appendix) matches the more precise simulation well. When we use this
exact numerical Python routine for the \emph{nuclear-recoil} yield, each call takes around
1\,minute. Since our fitting routines need to call this function thousands of times we use an
approximation that is higher order than the EDELWEISS approximation (so is more accurate, but not
exact) but has a smaller computation time, making it usable for our purposes (see
Sec.~\ref{sec:edw_sig_extract} and the Appendix).

\section{\label{sec:edw_sig_extract}Established germanium ionization yield width}
In the EDELWEISS publication~\cite{MARTINEAU2004426} it is clear that the ionization yield
width of nuclear-recoil events is systematically larger than expected. It is our goal to use a
fitting technique to quantify precisely how much larger the measured ionization yield width for
the EDELWEISS GGA3 detector is than expected (see Eq.~\ref{eq:sigma_Q_EDW} for the expectation) as
a function of recoil energy.

It has been noted that the expected ionization yield width for nuclear recoils given in
Eq.~\ref{eq:sigma_Q_EDW} is derived from a lowest-order ``moment expansion'' of the definition of
the ionization yield random variable, $Q$. While this approximation is not bad for the
electron-recoil ionization yield (see Fig.~\ref{fig:edw_ERQ}), it is not as accurate for the
nuclear recoil version because of the smallness of $\bar{Q}$. For that reason a moment expansion
out to order $1/E_r^6$--denoted by \signrI--is used in our fitting for both the electron and
nuclear recoil ionization yield width functions (see the Appendix for details). 

In the EDELWEISS publication~\cite{MARTINEAU2004426} the following functional form for the
ionization yield is used because it fits the mean of the ionization data well:
\begin{equation}\label{eq:mean_Q}
	\bar{Q} = AE_r^{B}.
\end{equation}
We adopt this form of the average ionization yield in order to extract the ``additional''
ionization yield width. EDELWEISS has extracted this additional yield width by assuming a
constant, called $C$, needs to be added in quadrature to the result of Eq.~\ref{eq:sigma_Q_EDW}
and using the measured ionization yield widths to fit for the value of that parameter. We execute
a similar fit, using the EDELWEISS measured points for the detector GGA3, and the corresponding
resolutions but with a slightly more flexible function that allows $C$ to be a function of the
recoil energy: $C(E_r) = C_0 + m\cdot E_r$, with $C_0$ and $m$ parameters. In our fit, the more
exact curve for the expectation of the ionization yield width (derived from
Eq.~\ref{eq:erq_joint}) is used. We use a Markov-Chain Monte Carlo (MCMC)
technique~\cite{Foreman_Mackey_2013} to be sure to populate the full posterior distribution in the
parameter space and account properly for parameter correlations.  To incorporate systematics, the
fit is taken over a six-dimensional space: $C_0$, $m$, $A$, $B$, $a_H$, and $\eta$. The last
variable is a fractional multiplier applied to the detector voltage to account for possible
measurement deviations in that detector setting. 

\begin{figure}[!h]
      \includegraphics[width=\columnwidth]{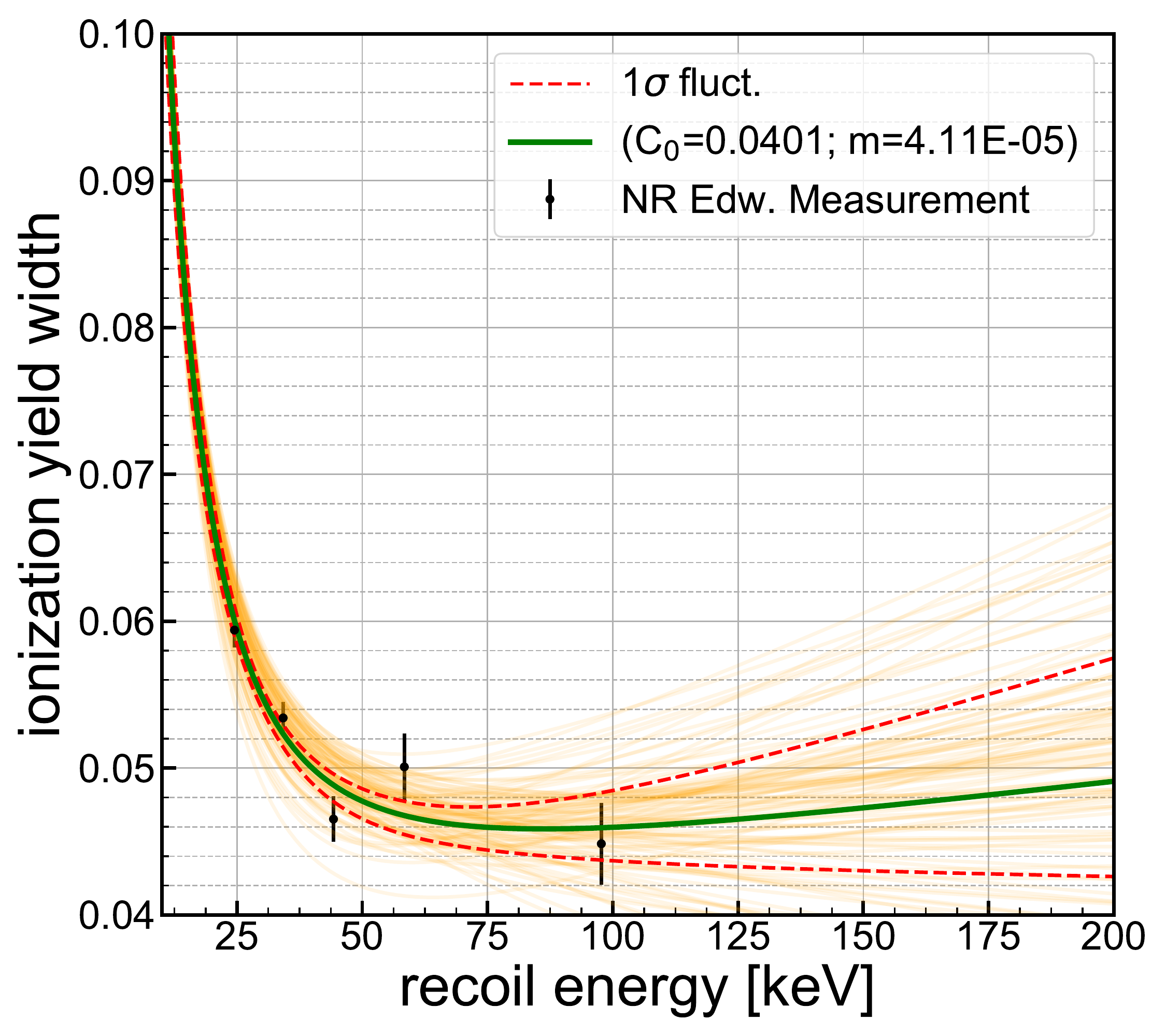}

	\caption{\label{fig:edw_C} (Color online) Our fit to the nuclear-recoil ionization width
	using the MCMC procedure. The solid curve is the maximum-likelihood fit to the $C$
	function, and the dashed lines are the assessed 1$\sigma$ statistical uncertainty bounds.
	The data points are the EDELWEISS~\cite{MARTINEAU2004426} measured values for detector
	GGA3 and the transparent curves are a sampling of 100 realizations of $C(E_r)$ using
	parameters pulled from the posterior parameter distributions.  }
\end{figure}

The result of the fit is shown in Fig.~\ref{fig:edw_C} with the maximum likelihood curve for the
extracted nuclear-recoil yield band width and several randomly-sampled curves from the correct posterior
distribution.  For the full reproducible code for this fit see the public data
release~\cite{Villano_Roberts_2022}.


The nuclear-recoil band width is well-reproduced by using the fitted $C(E_r)$ added in quadrature
to the base-level estimate. The base-level estimate is given in the Appendix and is symbolically
referred to as \signrI. Given the flexibility of our exact model for the ionization yield
distribution, we proceeded to use this $C(E_r)$ and its associated error to obtain the variance on
$N$ as a function of recoil energy, parameterized by the intrinsic nuclear recoil Fano factor, \fnr.

\section{\label{sec:ms_corr}Multiple scattering correction}
For the EDELWEISS data the nuclear-recoil ionization yield information is generated via scattering
of neutrons from a $^{252}$Cf source. Because of the use of neutrons, multiple scattering is an
obvious effect that will increase the measured ionization yield width. The EDELWEISS
study~\cite{MARTINEAU2004426} accounted for this effect using a Monte Carlo simulation and
concluded:

\begin{quote}
Although multiple interactions tend to lower $\langle Q \rangle$; this effect remains weak, and
the $Q$ distribution associated with single interactions events is only slightly narrower and
completely included in the wider band.
\end{quote}
the ``wider band'' being the band that encompasses the measured nuclear recoils. In other words
multiple scattering can \emph{not} account for the full observed ionization yield width. 

We have re-simulated the effect of multiple scattering in a detector that matches the EDELWEISS
GGA3 germanium detector (approx. cylindrical with a 70\,mm diameter and 20\,mm thick). For this
we have used a \texttt{Geant4}~\cite{1610988,AGOSTINELLI2003250} simulation where the
geometry--aside from the germanium detector--was not made identical to the EDELWEISS setup, but
where generic elements like typical cryostat materials and polyethylene shielding were included.
Our specific geometry (from inside to outside) included: the germanium detector; an electronics
``tower'' made mostly of copper with small amounts of insulating carbon; an ``inner vacuum
chamber'' wall made of stainless steel; liquid helium; a stainless steel Dewar with vacuum jacket;
and a rectangular polyethylene shield and supporting structure (aluminum). The source is located
between the Dewar and polyethylene shield, 66\,cm below the detector at a radial distance of
35\,cm from the cylindrical axis of the Dewar and germanium detector. 

Our simulation uses Geant \texttt{4.10.1.p02} and the so-called ``Shielding'' physics
list~\cite{g4physlist}. The main attribute of this physics list in the context of our analysis is
the high-precision neutron-scattering library for neutron energies below 20\,MeV. The use of this
``\texttt{NeutronHP}'' library~\cite{g4physref,THULLIEZ2022166187} gives more precise realizations
of the nuclear recoils because of the implementation of the detailed low-energy neutron
interaction library \texttt{G4NDL}. A small drawback of the library is that it sacrifices strict
energy-momentum conservation on an event-by-event basis, but that is not an important deterrent
for this study since the recoil spectrum is more correct. 

While the simulation setup does not match the EDELWEISS geometry, we point out that the geometry
will principally affect the energy distribution of the neutron flux near the detector. The yield
width is insensitive to that distribution because all of our scattering neutrons lie above 20\,keV,
where the elastic scattering cross section is away from the resonance region, relatively flat, and
well known~\cite{refId0}. Therefore, the distribution of multiple scatters \emph{within} the
detector--which does affect the $Q$ distribution--will not depend strongly on the energy
distribution of the neutron flux or the geometry, but rather if the germanium elastic cross
sections used are close to reality. The elastic cross sections used in our version of
``\texttt{NeutronHP}'' are in an energy region that has been well measured and match other
evaluations like the JENDL\,5.0 evaluation~\cite{refId0}.

We use this simulated data by applying the ionization yield model used in Sec.~\ref{sec:edw_yield}.
More precisely, we ``tune'' the sensor resolutions in the same way as produced the best match to
the electron-recoil band width, take the ionization yield to be $\bar{Q} = 0.16E_r^{0.18}$, and
take the intrinsic nuclear recoil Fano factor to be zero.

\begin{figure}[!h]
    \includegraphics[width=\columnwidth]{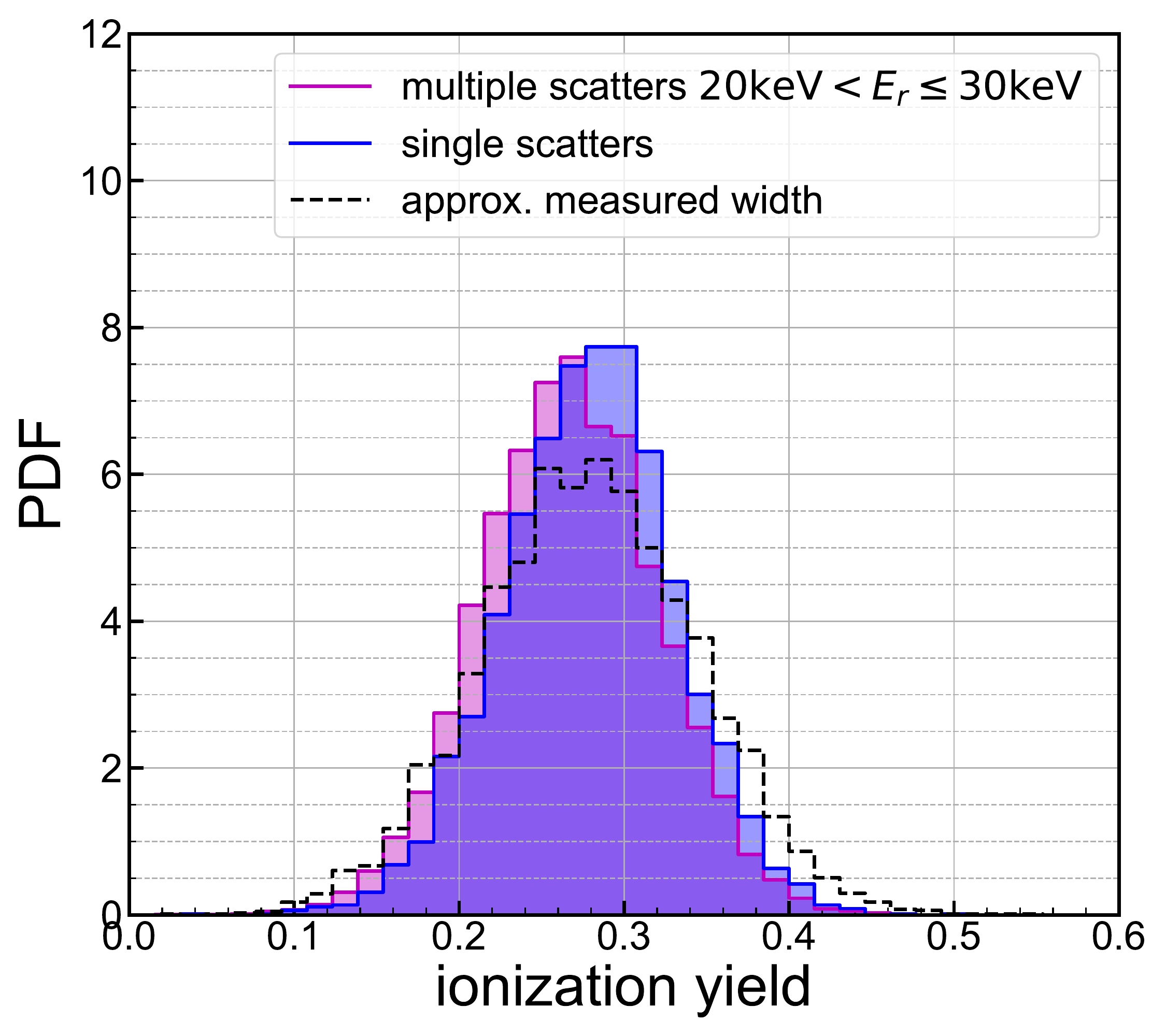}
    \caption{\label{fig:ms_hist} (Color online) Simulated ionization yield histograms of the
	single-scatter distribution (blue) and all-scatter distribution (magenta) for the energy
	range 20--30\,keV. The black dashed histogram are single scatters with an ``extra''
	resolution applied to bring the width of the distribution in line with what was measured
	by EDELWEISS as discussed in their publication~\cite{MARTINEAU2004426}.
    }
\end{figure}

The simulated ionization yield distributions in Fig.~\ref{fig:ms_hist} show that the single
scatter contribution has a clearly higher average yield than the distribution that includes all
scatters. However, the width of the distribution is only modestly wider over the energy range
shown (20--30\,keV). The empirical distribution (Fig.~\ref{fig:ms_hist} black dashed histogram) is
clearly significantly wider than our simulation with multiple scatters included--a feature that
gets more significant with increasing energy. 

\begin{figure}[!h]
    \includegraphics[width=\columnwidth]{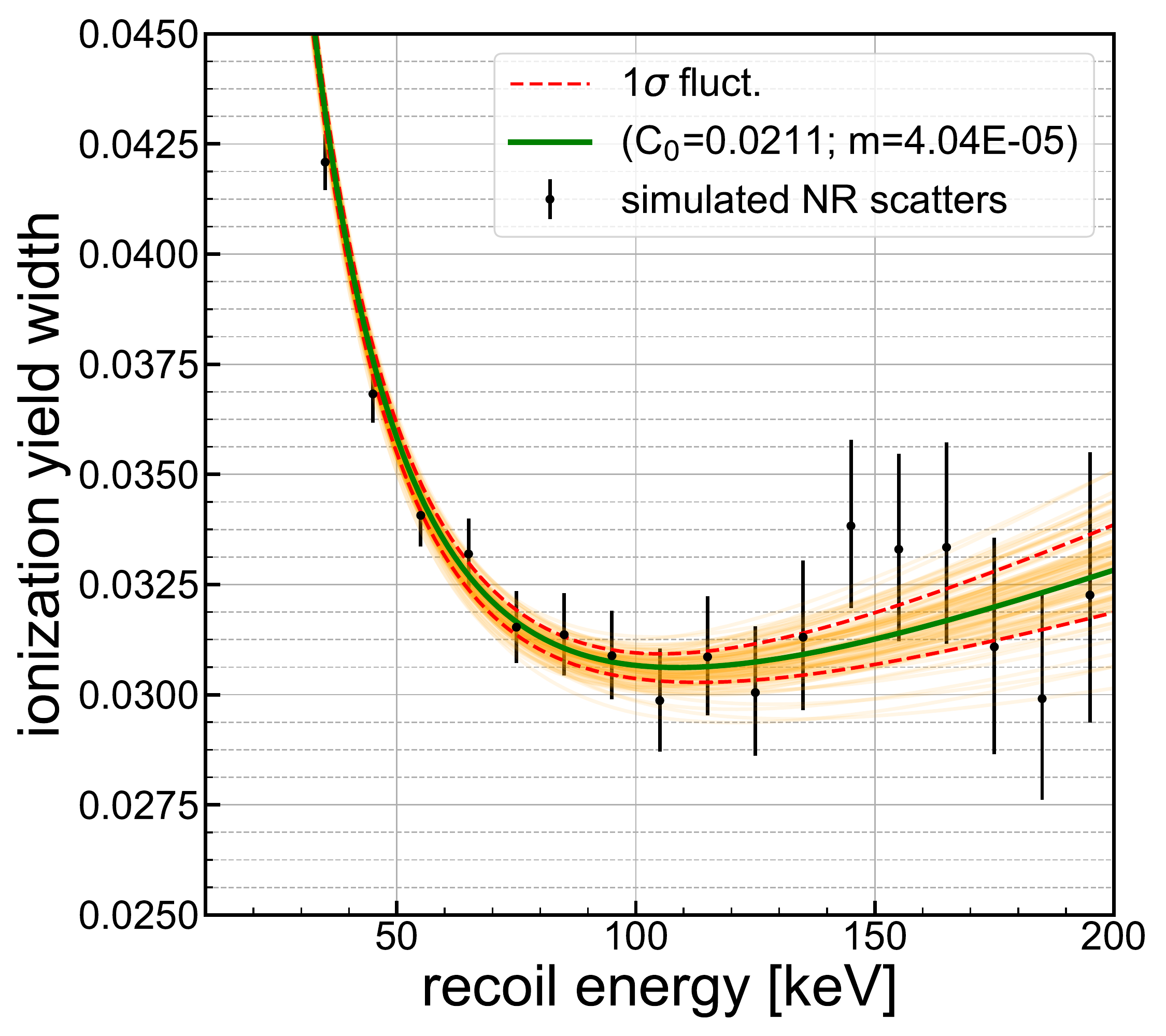}
    \caption{\label{fig:ms_corr} (Color online) A fit to the simulated multiple-scatter ionization
	width using the \texttt{Geant4} recoil data, and our yield model. The points are the ionization
	yield widths of the simulated data, the solid line is the maximum likelihood fit to the
	$C_m$ function. The dashed lines are the assessed 1$\sigma$ statistical uncertainty
	bounds. The transparent curves are a sampling of 100 realizations of $C_m(E_r)$ using
	parameters pulled from the posterior parameter distributions.
    }
\end{figure}

We have systematically fit the distribution widths from the simulation as a function of energy and
compared them with the single-scatter width predictions discussed previously.
Figure~\ref{fig:ms_corr} shows the ionization widths that result for a full \emph{simulated}
$^{252}$Cf data set with multiple scattering included. Of course, the resulting ionization widths
are larger than would result from a nuclear recoil sample consisting only of single scatters.
Since our ionization yield model only makes predictions for single scatters we compare the multiple
scatters to that prediction to see how much wider the ionization yield distribution becomes.  As
in the previous section we fit a function $C_m(E_r) = C_{0m} + m_m E_r$ that describes the
quadrature addition necessary to bring the single-scatter prediction in line with the simulated
multiple-scatter results. In this case we do not let $A$, $B$, $a_H$, or $\eta$ vary but set them
equal to their best fit values from the MCMC in Sec.~\ref{sec:edw_sig_extract}. The varying fit
parameters are $C_{0m}$ and $m_m$. 

It is clear from the fit displayed in Fig.~\ref{fig:ms_corr} that the quadrature addition needed
to describe the effect of multiple scattering is observable but significantly \emph{less} than what
is required to describe the EDELWEISS ionization yield width data. This multiple-scatter correction
to the yield widths will be used in Sec.~\ref{sec:extraction} to extract the required
\emph{additional} correction needed to describe the EDELWEISS data. We argue that this additional
correction is related to unaccounted uncertainty in the fundamental ionization production by
nuclear recoils; and can be described by an intrinsic nuclear recoil Fano factor. 

\section{\label{sec:extraction}Extracting the germanium intrinsic Fano factor}
We posit that the reason the measured ionization variance on EDELWEISS' GGA3 detector is larger
than the expected when including multiple scattering (see Sec.~\ref{sec:ms_corr}) is an
unaccounted intrinsic ionization variance in the nuclear scattering process. We quantify this
additional variance, by taking the quadrature subtraction of the corrections extracted in
Secs.~\ref{sec:edw_yield} and~\ref{sec:ms_corr}. The result is a correction, $C^{\prime}(E_r)$,
that is equal to the intrinsic ionization variance. Equation~\ref{eq:c_eq} shows the relationship
of the intrinsic variance to the previous corrections.

\begin{equation}\label{eq:c_eq}
	C^{\prime}(E_r) = \sqrt{C(E_r)^2 - C_m(E_r)^2}
\end{equation}

\begin{figure*}[!htb]
    \includegraphics[width=2.2\columnwidth]{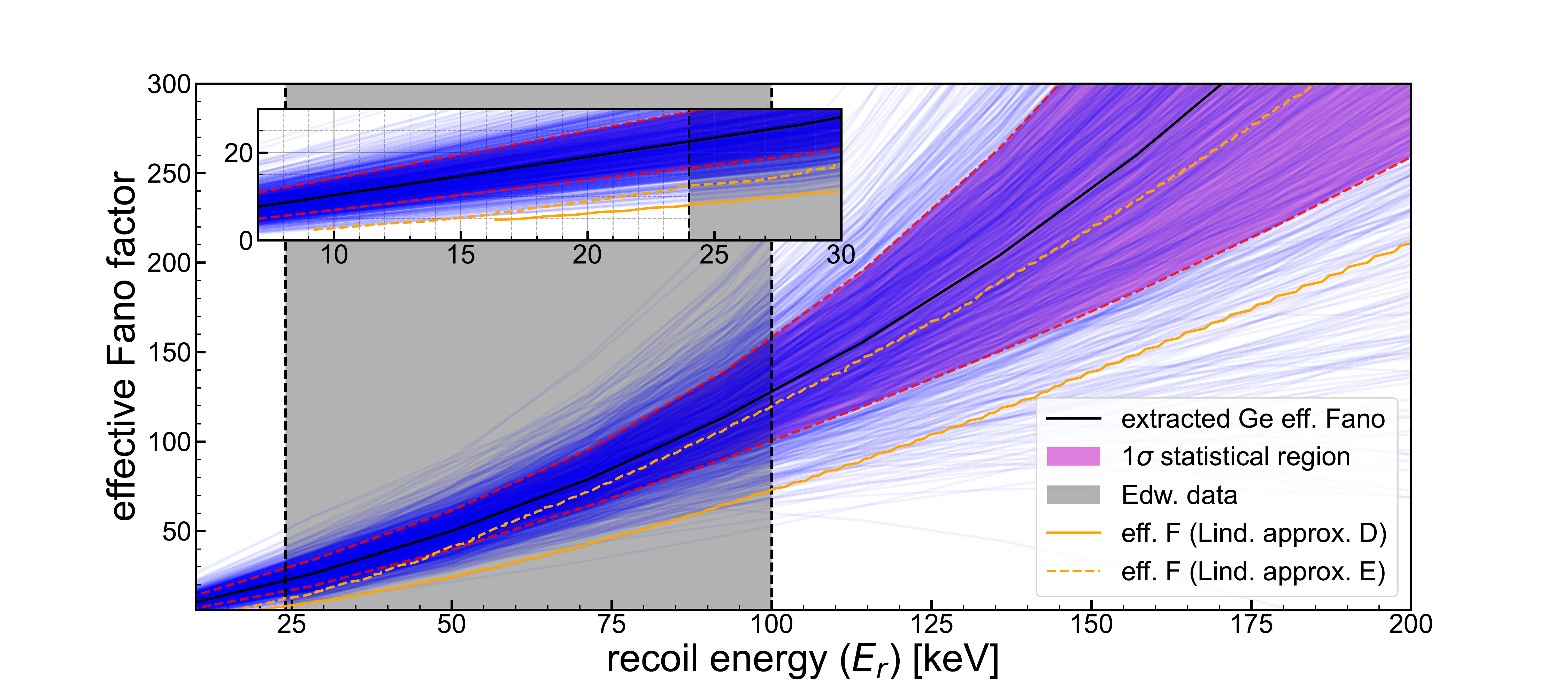}
    \caption{\label{fig:germanium_effF} (Color online) The extracted nuclear-recoil intrinsic Fano factor for
	germanium using the EDELWEISS~\cite{MARTINEAU2004426} data. The black line is the best
	fit, the magenta shaded region denotes
	the 1$\sigma$ statistical uncertainty region, and the inset shows a zoom of the region
	between 7 and 30\,keV. The blue transparent curves are 1000 samples drawn from the MCMC
	posterior distribution--these give a sense for the presence of outlying behaviors. The
	gray shaded region is the region in which there are EDELWEISS yield width data; in that
	region the results are \emph{not} an extrapolation. The Lindhard prediction for the
nuclear recoil Fano factor are given in orange for the Approximations D (solid) and E
(dashed). To produce those predictions we used a Lindhard model for the ionization yield with
parameter $k$=0.157~\cite{PhysRevD.105.122002}. 
	}
\end{figure*}

Our intrinsic ionization variance is then converted into a Fano factor for nuclear recoils, \fnr,
as advocated in Sec.~\ref{sec:si_fano}. The conversion to the nuclear recoil Fano factor is made
by assuming the intrinsic variance is produced by simply increasing the nuclear recoil Fano factor
from \fnr=0 to some finite (positive) value within the framework of the model given in
Eq.~\ref{eq:erq_joint} by setting the variance on the independent random variable $N$ taken to be
$\sigma_N = \sqrt{\fnr \bar{N}}$. The actual value of $\fnr(E_r)$ is then simply given by:

\begin{equation}\label{eq:Fn_def}
 \signrI(E_r;\fnr) = \sqrt{C^{\prime}(E_r)^2 + \signrI(E_r;\fnr=0)^2}
\end{equation}
Figure~\ref{fig:germanium_effF} shows the extracted intrinsic Fano factor, \fnr\,, as a function of
the recoil energy.
The estimate for the uncertainties on the resulting $\fnr(E_r)$ were obtained from the MCMC
posterior distribution of all of the parameters ($A$,$B$,$a_H$,$\eta$,$C_0$,$m$) in the original
fit and the posterior distribution of the $C_{0m}$ and $m_m$ parameters in the multiples fit. A
single realization of $\fnr(E_r)$ is obtained by using a sample of the original and multiples fit
and then subtracting them in quadrature to get $C^{\prime}(E_r)$. Each sample of $C^{\prime}$ is
turned into a sample of \fnr\,through Eq.~\ref{eq:Fn_def}. The maximum likelihood parameters are
taken as the central value for \fnr\,and we obtain the approximate 1$\sigma$ deviations by taking
the standard deviation of all samples at each energy--these are plotted as the magenta band in
Fig.~\ref{fig:germanium_effF}. 

These uncertainties include the systematic uncertainty on the result with contributions from
several parameters which, while nominally fixed, are not known with certainty. They are, in order
of decreasing importance: multiple scattering; a finite-binning uncertainty on the EDELWEISS
ionization yield data; a possible departure of the quantity $V/\epsilon_{\gamma}$ from the nominal
4/3 value (fit parameter $\eta$); charge trapping (fit parameters $A$ and $B$); and the functional
form of the average ionization yield.  The uncertainties are obtained by directly estimating the
contribution (in the case of the finite-binning) or including nuisance parameters in the
6-parameter MCMC~\cite{Foreman_Mackey_2013} fit to the EDELWEISS ionization yield width data for
GGA3 for the extraction of $C(E_r)$. For each of the parameters representing the systematic
uncertainties, a prior was chosen that was reflective of the state of knowledge on the parameters. 
\begin{table}[!hbt]
\begin{tabular}{ c  c  c  c}
\hline
\hline
	Classification  &  Size (\%) & Param. & Relevant Corr.   \\ \hline
	statistical (U)  & 40-80  & $C_0$, $m$ & none \\
	multiple scattering (U)  & $<$\,6  & none & none \\
	finite-binning (U)  & 5  & -  & - \\
	$V/\epsilon_{\gamma}$ (U) & $<$\,20 & $\eta$ & $a_H$ \\
	charge trapping (U)  & $<$\,20 &  $A$, $B$ & $C_0$, $m$  \\
	yield variation (U)  & $<$\,20 &  $A$, $B$ & $C_0$, $m$ \\
	multiple scattering (C)  & 60-70  & none & none \\

\hline
\hline
\end{tabular}
   \caption{\label{tab:sys} The uncertainties and correction sizes for the extraction of the
	nuclear-recoil intrinsic Fano factor for germanium using the EDELWEISS~\cite{MARTINEAU2004426} data. The
	first column lists uncertainties with a (U) and corrections with a (C). The third column
	are the parameters in the fit related to that category (if any). The
	last column lists the fit parameters that have relevant correlations with that particular
	category.
	}
\end{table}
The total uncertainty is estimated in Figure~\ref{fig:germanium_effF} and the factional impact of
each of the uncertainties or corrections are given in Tbl.~\ref{tab:sys}.

\section{\label{sec:conclusions}Conclusions}
We have espoused the preference for quantifying the inherent uncertainty on the number of
electron-hole pairs produced as an intrinsic Fano factor, \fnr, for nuclear recoils. We have
also presented constraints on such a parameter from previous measurements on silicon and
germanium--two important target materials for precision low-mass dark matter
searches~\cite{PhysRevD.95.082002}. In the latter case we extracted meaningful \fnr~measurements
by a technique that can be adapted to low-threshold detectors measuring ionization and heat, but
that did not require a specialized neutron scattering setup. We have used the Lindhard predictions as a guide,
with the hope that future experiments will be able to distinguish between approximations in that
work and/or inspire the development of a more accurate framework. 

Our results indicate that the intrinsic nuclear-recoil Fano factor is larger than expected for
both silicon and germanium--24.3$\pm$0.2 and 26$\pm$8 respectively at 25\,keV recoil energy. The
expectation in some literature is based on the assumption that the number of phonons created is a
Poisson random variable~\cite{Mei_2020}. In that case the electron-recoil Fano factor is around
0.13 for germanium~\cite{PhysRevB.22.5565} and the intrinsic nuclear-recoil Fano factor should be
larger by about a factor of $1/\sqrt{\bar{Q}}$--still far lower than our suggested values. In the
authors' view, this would seem to indicate that for nuclear recoils the number of created phonons
is \emph{not} Poisson distributed and has a distribution that is significantly wider than naively
expected; this wider distribution could then be imprinted on the electron-hole pairs in a way
similar to the derivation in~\cite{Mei_2020}. The authors do not see any reason why the number of
phonons produced should have a Poisson distribution, in fact the Lindhard references explicitly
compute an ionization variance that are out of line with that assumption~\cite{osti_4701226}.
The Lindhard predictions for the intrinsic nuclear-recoil Fano factor are shown in
Fig.~\ref{fig:germanium_effF} for germanium and have an \fnr\,at least as large as 8 at 25\,keV
recoil energy. Those intrinsic nuclear-recoil Fano predictions are not inconsistent with our
measurement above around 50\,keV but appear to be systematically lower than our measurement below
50\,keV--perhaps due to an ionization yield that decreases more sharply toward lower recoil energy
than the Lindhard theory suggests. 

Based on our ionization yield model, which can describe EDELWEISS data well, the variance induced
by the intrinsic Fano factor is correlated in its effect on ionization and heat resolutions.
Roughly speaking, this means that the widening of ``nuclear recoil bands'' in low-threshold dark
matter searches with discrimination capabilities (like SuperCDMS~\cite{PhysRevLett.112.241302} and
EDELWEISS~\cite{Hehn2016}) may be smaller than one would naively expect. 

There is a lot of existing data that might be exploited using our technique, but it is often true
that precise resolution data is not published. If the sensor resolution is carefully extracted,
then our technique might serve to extract \fnr~more precisely for both silicon and germanium in
the low-energy region. Such information is invaluable to low-mass nuclear-recoil dark matter
searches in silicon and germanium that employ detectors \emph{without} nuclear-recoil
discrimination capabilities.  

\begin{acknowledgements}
The authors would like to acknowledge Brian Dogherty, Allison Kennedy, Danika MacDonell and members
SuperCDMS Collaboration for discussions on topics related to the ionization variance.  We would
also like to thank Kitty Harris for a careful reading of the manuscript. This work
was partially supported by DOE grant DE-SC0021364.  
\end{acknowledgements}

\appendix*
\section{\label{sec:appendix}Calculation of \signr and cross-checks}
\subsection{\label{subsec:signr}Calculation of \signr}
Both the EDELWEISS and SuperCDMS detectors can be correctly modeled by assuming the measurements
of the ionization and heat depend on three (approximately) independent random variables: the
number of electron-hole pairs created in a detectable interaction, $N$; the variation (noise
fluctuations) in the ionization sensor, $\delta I$; and the variation in the heat detection
$\delta H$.  The distributions of $\delta I$ and $\delta H$ have zero mean and are approximately
normally distributed with an energy-dependent standard deviation given by the ionization and heat
sensor resolutions. The typical measured quantities in these experiments are specific combinations
of those random variables defined thusly:

\begin{equation}\label{eq:var_defs}
	\begin{aligned}
		\tilde{E}_r &\equiv E_r + \left ( 1 + \frac{V}{\epsilon_{\gamma}} \right ) \delta H -
		\frac{V}{\epsilon_{\gamma}} \delta I \\
		Q &\equiv \frac{\epsilon_{\gamma} N + \delta I}{\tilde{E}_r}.
	\end{aligned}
\end{equation}

The variable $\tilde{E}_r$ is the \emph{measured} recoil energy, $Q$ is the measured ionization
efficiency (yield), $E_r$ is the \emph{true} recoil energy, and $V$ is the voltage across the
cylindrical detector. With this model, if the sensor resolutions are published (or otherwise known),
the only remaining things needed to predict exact distributions for all the measured quantities
are the true recoil energy distribution (which can be simulated) and the distribution of the
random variable $N$. The latter is directly related to the Fano factor or the intrinsic 
nuclear-recoil Fano factor. Since $N$ is rather high for recoil energies above $\sim$10\,keV the
distribution is taken to be approximately normal, with the mean given by the average ionization
yield at the particular recoil energy ($\bar{Q}(E_r)$) and the width being given by the intrinsic 
Fano factor, \fnr.

We have done the exact calculation simply by recognizing the joint conditional probability
distribution for $\tilde{E}_r$ and $Q$ must have the following form:

\begin{widetext}
\begin{equation}\label{eq:erq_joint}
\begin{aligned}
	P(Q,\tilde{E}_r | \delta H, \delta I, N, E_r) &= \delta\left(\tilde{E}_r - \left[E_r + \left(1+\frac{V}{\epsilon_{\gamma}}\right)\delta H -\left(\frac{V}{\epsilon_{\gamma}}\right)\delta I\right] \right) \\
	&\times \delta \left(Q - \left[\frac{\epsilon_{\gamma} N + \delta I}{E_r + \left(1+V/\epsilon_{\gamma}\right)\delta H - (V/\epsilon_{\gamma}) \delta I}\right]\right).
\end{aligned}
\end{equation}
\end{widetext}

Equation~\ref{eq:erq_joint} will correctly give the ionization yield ($Q$) distribution at a
single measured energy or over a range of measured energies. The distribution close to normal for
a wide range of parameters but not exactly normal. The distribution is especially far from normal
when the heat or ionization have a large enough variance so that the measured recoil energy
becomes consistent with zero. The ionization yield standard deviation with this ''exact''
calculation is referred to as $\signr$.

The procedure outlined above involves integrals that are difficult to accomplish analytically. For
that reason, slower numerical techniques are used and the computation time makes it difficult to
use (around 1\,min for one calculation at one energy and parameter-value point). In this work, as
discussed in Sections~\ref{sec:edw_yield},\ref{sec:ms_corr}, and~\ref{sec:extraction}, the fitting
requires many evaluations of the function and so it must be approximated.

Part of the problem is not only the functional dependence on $E_r$, but the functional dependence
on our nuisance parameters $A$, $B$, $a_H$, and $\eta$. In the general case--nuclear recoils
with average yield modeled by the $A$ and $B$ parameters--we compute the ``moment'' expansion
of $Q$ in Eq.~\ref{eq:var_defs} to order $1/E_r^6$. We refer to this expression as \signrIst.  For
electron recoils, the agreement is quite good if we simply take this expansion with $A$=1 and
$B$=0 (see Fig.~\ref{fig:edw_ERQ}).  The expansion to lower order ($1/E_r^2$) is the expression used
by EDELWEISS--\signrEDW (see Eq.~\ref{eq:sigma_Q_EDW}). 


For nuclear recoils, the agreement is not as good, so we add a correction based on the preferred
values of the nuisance parameters from our fit to the EDELWEISS data. Taking $A_0$=0.149,
$B_0$=0.178, $a_{H0}$=0.038, and $\eta_0$=1.000 we can use the exact function to create a static
correction to for use in the nuclear recoil case. This is the approximation we use to describe our
nuclear recoil ionization yield widths in our fitting procedure of Sec.~\ref{sec:edw_sig_extract}:

\begin{widetext}
\begin{equation}\label{eq:sigma_Q_NAI}
\begin{aligned}
	\signrI(E_r,A,B,\eta,\fnr=0)^2 &= \signrIst(E_r,A,B,\eta,\fnr=0)^2 \\
	&+ \left[\signr(E_r,A_0,B_0,\eta_0,\fnr=0)^2 - \signrIst(E_r,A_0,B_0,\eta_0,\fnr=0)^2
	\right].
\end{aligned}
\end{equation}
\end{widetext}

The form shown in Eq.~\ref{eq:sigma_Q_NAI} is much faster to compute than the exact version, but
gives ionization yield widths which differ from the exact model by at most 7\% over our parameter
space ($E_r$ plus nuisance parameters).

\subsection{\label{subsec:erfano}Cross check with electron recoil Fano factor}
One excellent check for consistency of our method is to fit the electron recoil ionization yield
band and extract the electron recoil Fano factor, F. We cannot accomplish this with the real
EDELWEISS data~\cite{MARTINEAU2004426} that we've used for the majority of this paper because the
data is not precise enough (about 10\% relative uncertainty) and the Fano contribution to the
ionization yield variance is expected only to be around 0.1\%. This is in contrast to the nuclear
recoil Fano contribution which we have measured to be at least 10\%.  

Instead what we've done is simulate electron recoil band data in a similar way as was done in
Fig.~\ref{fig:edw_ERQ}, the high-precision simulated data points. We selected a Fano factor of
F=0.15, adjusted the instrumental resolution so that the Fano contribution was about 1\% in yield
variance, and simulated the data with about 0.5\% relative uncertainty on each data point. With
this fit--using the same MCMC fitting method we used for nuclear recoils in this work--we
extracted a Fano factor of F=0.13$\pm$0.08, consistent with the Fano factor we set for the
simulation (F=0.15). 

\clearpage
\bibliography{prdrefs_short.bib}
\bibliographystyle{apsrev4-1}

\end{document}